\documentclass[aps,superscriptaddress,eqsecnum,nofootinbib,showpacs,preprintnumbers]{revtex4}
\usepackage{graphicx,epsfig}
\usepackage{amsmath}
\usepackage {amssymb}
\usepackage{color}

\newcommand{\be}{\begin{eqnarray}}
\newcommand{\ee}{\end{eqnarray}}
\newcommand{\bea}{\begin{eqnarray}}
\newcommand{\nn}{\nonumber}
\newcommand{\eea}{\end{eqnarray}}

\def\de{\partial}
\def\a{\alpha}
\def\b{\beta}
\def\g{\gamma}

\def\d{\delta}
\def\D{\Delta}
\def\e{\eta}
\def\la{\lambda}
\def\La{\Lambda}
\def\k{\kappa}
\def\m{\mu}
\def\n{\nu}
\def\r{\rho}

\def\s{\sigma}

\def\p{\pi}

\def\z{\zeta}

\def\ie{{\it i.e. }}

\begin{document}

\title{Anisotropic Extra Dimensions}

\author{Eleftherios Papantonopoulos}

\author{Antonios Papazoglou}

\author{Minas Tsoukalas}

\affiliation{Department of Physics, National Technical University
of Athens, Zografou Campus GR 157 73, Athens, Greece}

\date{\today}

\begin{abstract}

We consider the scenario where  in a five-dimensional theory, the
extra spatial dimension has different scaling than the other four
dimensions. We find background maximally symmetric solutions, when
the bulk is filled with a cosmological constant and at the same
time it has a three-brane embedded in it. These background
solutions are reminiscent of Randall-Sundrum warped metrics, with
bulk curvature depending on the  parameters of the breaking of
diffeomorphism invariance.  Subsequently, we consider the scalar
perturbation sector of the theory and show that it has certain
pathologies and the striking feature that in the limit where the
diffeomorphism invariance is restored, there remain ghost
scalar mode(s) in the spectrum.

\end{abstract}

\maketitle

\section{Introduction}

Theories with extra spacetime dimensions were introduced in the
'20s  by the work of  Kaluza and Klein (KK) and gained further
attention with the advance of string theory, where they were a
crucial ingredient for the self-consistency of the theory. During
the last decade, the interest in model building and phenomenology
of extra dimensional theories has been renewed with the discovery
that they can play an important role in physics of
energies/distances which are probed in current experiments and
observations \cite{AADD}. There have been a wealth of models were
extra dimensions open up at the electroweak scale and provide new
insights of standard high energy problems \cite{randall}, or  at
macroscopic scales and modify gravity in the infrared
\cite{Dvali:2000hr}.

A common factor of these models is that the starting point in
model building is a higher dimensional diffeomorphism invariant
theory. The background solutions of the metric,  \ie the
gravitational vacuum of the theory, spontaneously breaks the
higher-dimensional diffeomorphism group down to its
four-dimensional subgroup. This constitutes a gravitational Higgs
mechanism and provides masses for the towers of KK gravitons,
vectors and scalars \cite{Cho:1992rq,Kakushadze:2000zn,Kakushadze:2007dj}.

Recently, Ho\v rava proposed a theory of gravity which also breaks
diffeomorphism invariance \cite{Horava:2009uw}. The new theory is
supposed to be an adequate ultraviolet (UV) completion of General
Relativity above  the Planck scale. Its basic assumption is the
existence of a preferred foliation by three-dimensional constant
time hypersurfaces, which splits spacetime into space and time.
This allows to add higher order spatial derivatives of the metric
to the action, without introducing higher order time derivatives.
This is supposed to improve the UV  behaviour of the graviton
propagator and render the theory power-counting renormalisable
without introducing ghost modes, which are common when adding
higher order curvature invariants to the action in a covariant
manner \cite{Stelle:1977ry}.

Such a theory cannot be invariant under the full set of
diffeomorphisms, but it can still be invariant under the more
limiting foliation-preserving  diffeomorphisms. This breaking of
the full diffeomorphisms group down to its foliation-preserving
subgroup is, however, explicit and not spontaneous. This has been
the source of problems, related with a scalar degree of freedom,
whose behaviour   creates a serious viability issue for all
versions of the theory. In the version with ``projectability'' and
for maximally symmetric backgrounds, the scalar is either
(classically) unstable  or it becomes a ghost
(quantum-mechanically unstable)
  \cite{Sotiriou:2009bx,Charmousis:2009tc,Koyama:2009hc,Bogdanos:2009uj} \footnote{See however
\cite{Mukohyama:2010xz} and references therein for conditions
under which the  classical instability does not show up.}. Also
the
  non-projectable version of the theory has strong coupling problems and
  instabilities \cite{Blas:2009yd,Papazoglou:2009fj,Padilla:2010ge}.

In the present paper, we will revisit the extra dimensional
theories and break explicitly, \ie at the action level, the higher
dimensional diffeomorphism invariance to its foliation-preserving
subgroup. The foliation that we will choose is adapted to an extra
space dimension, as opposed to Ho\v rava-type of theories where it
was adapted to the time dimension. This will leave intact the
four-dimensional spacetime diffeomorphisms contrary to Ho\v rava
theory. It is evident that in the resulting theory we can not
demand the theory to be renormalisable, as it was the motivation
in the Ho\v rava theory, since that would indicate the inclusion
of  higher order {\it four-dimensional curvature} invariants, thus
introducing ghosts. On the other hand, depending on the difference
in scaling of the extra dimension compared with the scaling of the
four ordinary spacetime dimensions, higher powers of the extrinsic
curvature of the four dimensional hypersurfaces may play part in
the effective theory, before reaching energies where inevitably
ghost-bearing higher order dimensional operators appear.

The aim of this paper is to study such a theory where
diffeomorphism invariance in an extra dimensional model is
explicitly broken. We will pursue this aim in a  theory, which in
the limit of diffeomorphism invariance restoration, tends to the
Randall Sundrum warped model with one brane and infinite extra
dimension. We will formulate the theory, find background solutions
and study the scalar sector  perturbations around this background.
This study will reveal that such an explicit breaking of the
five-dimensional diffeomorphism invariance is dangerous and
introduces ghost scalar mode(s).

  The pathological behaviour of gravity at small distances  is a problem that  plunges
  many gravity theories. In the direction of proposing a UV-completion theory of General Relativity,
   higher order curvature terms
   were introduced. The  so-called
modified gravity theory, includes $f(R)$ gravity models \cite{fR},
and the Gauss-Bonnet (GB) models \cite{GB}. The $f(R)$ models have
a better behaviour in the UV but they have serious cosmological
problems \cite{Gano2} (see however \cite{Nojiri:2010wj}). For GB
models it was shown that tensor perturbations are typically
plagued by instabilities in the UV.

Similar problems arise by considering
modifications of gravity at large distances.
 The  Dvali-Gabadadze-Porrati (DGP) braneworld
model~\cite{Dvali:2000hr} and its extension~\cite{kofi} with a GB
term in the bulk, was proposed as an alternative to the
acceleration of the Universe without the need of dark energy
\cite{Sahni,Lue}. However, it was soon realized that  the DGP
model contains a ghost mode \cite{Koyama}, which casts doubts on
the viability of the self-accelerating solution. Other theories that modify gravity at large distances were proposed
by
 introducing  a
mass term to the graviton via a potential term. This was explored
many years ago by Fierz and Pauli \cite{Fierz}. Later it was
realized that gravitational theories with a mass term are behaving
quite unlike other theories and are typically accompanied by
several problems, most notably ghost/tachyon instabilities and
strong coupling issues. Also massive gravity theories have a very
characteristic feature the Van Dam-Veltman-Zakharov (vDVZ)
discontinuity \cite{vanDam:1970vg,Zakharov:1970cc}, \ie the fact
that as the mass of the graviton goes to zero the scalar graviton
mode fails to decouple (see, however, a way out in curved space
\cite{Kogan:2000uy,Porrati:2000cp}). To cure this problem one has
to advocate nonlinear dynamics, employing the Vainshtein mechanism
\cite{Vainshtein:1972sx}, but then a ghost appears in the spectrum
first noticed by Boulware and Deser \cite{Boulware:1973my}.

A similar behaviour like the vDVZ discontinuity in massive gravity
is observed in our theory: although the background solutions tend
to the Randall-Sundrum solutions as the diffeomorphism invariance
is restored, the problematic scalar modes are not removed from the
spectrum in the same limit. We attribute this behaviour to the
explicit breaking of the diffeomorphism invariance in the theory.

\section{Breaking the diffeomorphism invariance}

Our starting point is five-dimensional Einstein  gravity in the
presence of a cosmological constant and a brane embedded at some
point $y=0$ of the extra dimension.  We assume $Z_2$ symmetry
along the $y$-direction with $y=0$ as the fixed point.  The
corresponding action is given by
 \be\label{5daction1}
 S=\int d^5x \left(R^{\left(5\right)}-2\La_{5}\right)\sqrt{-G}-\int_{brane}d^4x
 \sqrt{-g}\,\s~,
\ee where $R^{\left(5\right)}$ is the five-dimensional
Einstein-Hilbert term and $\La_{5}$, $\s$ and $G_{MN}$ is the
five-dimensional cosmological constant, the brane tension and the
metric respectively. In the following we wish to modify the above
theory in a way that diffeomorphism invariance is broken along the
extra dimension.

First, let us make the following ADM splitting along the extra dimension $y$
 \be
ds^{2}=dy^{2} N^{2} c^{2}+g_{\m\n} \left(N^{\m}dy+dx^{\m}\right)
\left(N^{\n}dy+dx^{\n}\right)~, \ee with $g_{\m\n}$ the
four-dimensional metric. In the modification that we will discuss,
we shall adopt an anisotropic scaling of the different dimensions
with $\left[x^{\m}\right]=-1$ and $\left[y\right]=-w$, where the
extra dimension singles out. For such a theory, the ADM metric
components scale as $\left[g_{\m\n}\right]=0$, $\left[N\right]=0$
and $\left[N^{\n}\right]=w-1$.   The five-dimensional action
(\ref{5daction1}) can be generalised  as
 \be\label{5daction2}
 S=\int d^4 x dy N \sqrt{-g}\, \left[\frac{\r}{2}\left(R^{\left(4\right)}-2\La_{5}\right)-\frac{2}{\k^{2}} \left(K_{\m\n}
  K^{\m\n}-\la K^{2}\right)\right]-\int_{brane}d^4x\sqrt{-g}\,\s~,\ee
where the extrinsic curvature $K_{\m\n}$ is given by
$\left[K_{\m\n}\right]=w$, with \be K_{\m\n}=\frac{1}{2
N}\left(\partial_{y}
g_{\m\n}-\nabla_{\m}N_{\n}-\nabla_{\n}N_{\m}\right)~,\ee and the
scaling of the various quantities are $\left[\r\right]=w+2$,
$[\k]=\frac{w-4}{2}$,
$\left[R^{\left(4\right)}\right]=\left[\La_{5}\right]=2$. The
above action receives radiative  corrections and can be extended
by including higher dimensional operators. Depending on the value
of $w$, there can be higher powers of the extrinsic curvature
important at energies lower than the one that  unitarity of the
theory is lost (typically when $R^2$ terms dominate). We will
discuss about that possibility in a later section. For the moment
we will restrict ourselves to the classical generalised action
(\ref{5daction2}).

The parameter $\la$ is crucial for the following discussion. It
represents the  breaking of the five-dimensional diffeomorphism
invariance to its four dimensional subgroup. The restricted
symmetry allows for a different weighting of the $K_{\m\n}^2$ and
$K^2$ terms in the action, contrary to the fully covariant theory.

 The field equations coming from variation of (\ref{5daction2}) with respect to the fields $N,N_{\m}$ and $g_{\m\n}$ are given by

\bea
0=\frac{1}{\sqrt{-g}} \frac{\d S}{\d N_{\phantom{\m}}}&=&\frac{\r}{2}\left(R^{\left(4\right)}-2\La_{5}
\right)+\frac{2}{\k^{2}} \left(K_{\m\n} K^{\m\n}-\la K^{2}\right) \label{dNeq}~, \\
0=\frac{1}{\sqrt{-g}} \frac{\d S}{\d N_{\m}}&=&-\frac{4}{\k^{2}}\nabla_{\m} \p^{\m\n} \label{dNmeq}~, \\
0=\frac{1}{\sqrt{-g}} \frac{\d S}{\d
g_{\m\n}}&=&-\frac{\r}{2}\left(G_{\left(4\right)}^{\m\n}+g^{\m\n}\La_{5}\right)
N \label{dgmneq} \\ \nn
&&+\frac{2}{\k^{2}}\left[\partial_{y}\p^{\m\n}+NK\p^{\m\n}+2\nabla_{\s}\left(\p^{\s(\m}N^{\n)}\right)-N_{\k}\nabla^{\k}\p^{\m\n}
+2NK^{\s\m}\p^{\n}_{\,\,\s}\right] \\ \nn && -\frac{1}{2}g^{\m\n}
\frac{2}{\k^{2}} \left(K_{\m\n} K^{\m\n}-\la
K^{2}\right)-\frac{1}{2}\s \d\left(y\right)g^{\m\n}~, \eea where
the  $y$-canonical momentum of the four-dimensional metric is \be
\p^{\m\n}=K^{\m\n}-\la K g^{\m\n}~. \ee

As in the original case of Randall-Sundrum, it is important to
consider the junction conditions of the system. These junction
conditions involvng the brane tension, are found by identifying
the distributional terms in the equations of motion. It is fairly
easy to see from (\ref{dgmneq}), that the only distributional term
is $\partial_{y}\pi^{\m\n}$. Integrating along the extra dimension
and taking the limit close to the brane, gives the junction
conditions, which takes the following familiar form \be
\label{junc1}
\frac{2}{\kappa^{2}}\left[\pi^{\m\n}\right]^{+}_{-}=\frac{1}{2}g^{\m\n}\sigma~.
\ee Under the $Z_{2}$ symmetry the above relation can be rewritten
as \be \label{junc2}
\frac{4}{\kappa^{2}}\pi^{\m\n}=\frac{1}{2}g^{\m\n}\sigma~. \ee

In the following section, we will present background solutions for
this theory  which possess maximal symmetry in four dimensions. We
will see that these solutions bear great similarity with the
Randall-Sundrum solutions \cite{randall} and their curved version
\cite{Kaloper:1999sm}, with $\la$-dependent bulk curvatures.

\section{Solutions with Maximally symmetric backgrounds}

In this section we will consider  solutions of the field equations
where the four-dimensional metric $g_{\m\n}$ is maximally
symmetric. This metric evaluated at the brane position, is also
the induced metric on the brane at $y=0$, since the brane is
static. Therefore, the solutions for the bulk metric,  satisfying
the appropriate boundary conditions, define families of maximally
symmetric brane solutions.

\subsection{Flat Brane}

Let us first examine the case where $g_{\m\n}$ is flat. In  this
case we will consider the following ansatz for the metric
components \be \label{flatbrane}
N_{\m}=0,\,\,\,g_{\m\n}=e^{2\,f\left(y\right)}\e_{\m\n},\,\,\,N=1~,
\ee allowing for a warp factor $f(y)$ along the extra dimension.
Then, eq. (\ref{dNeq}) becomes

\be
\k\La_{5}\r+\frac{8\left(-1+4\la\right)\,\left(f'\left(y\right)\right)^{2}}{\k}=0
\label{dNeq1}~,  \ee and eq. (\ref{dNmeq}) is automatically
satisfied while  eq. (\ref{dgmneq}) yields

\be
 \k^{2} \La_{5} \r+\k^{2} \s \d\left(y\right) +8 \left(-1+4\la\right) \left(f'\left(y\right)\right)^{2}+4
 \left(-1+4\la\right)f''\left(y\right)=0~. \label{dgmneq1}
\ee

Depending on the value of the parameter $\la$ we distinguish the three following cases of solutions that we need to examine.

 \subsubsection{\underline{Case 1 $(\la<\frac{1}{4})$}}

 In this case, (\ref{dNeq1}) gives

\be \label{warpfact1} f\left(y\right)= -\frac{|y|\, \k}{2
\sqrt{2}} \sqrt{\frac{\La_{5} \,\r}{1-4 \la}}~. \ee Clearly, $\r
\,\La_{5}$ must be positive. Substituting the above solution to
 (\ref{dgmneq1}) we see that it is trivially
satisfied.

\subsubsection{\underline{Case 2 $(\la>\frac{1}{4})$}}

In this case, we have the same form of solution, but now
$\r\,\La_{5}$ is negative.   Again, eq. (\ref{dgmneq1}) is
satisfied. For $\la=1$ we get a Randall-Sundrum-like solution
\cite{randall}.

\subsubsection{\underline{Case 3 $(\la=\frac{1}{4})$}}

 For this case, we get that $\La_{5}=0$ and $f\left(y\right)$ is arbitrary. This is a critical point of the theory,
   where the bulk theory is conformally invariant.

 Substituting $g_{\m\n}$ in (\ref{junc2}) we get \be \kappa^{2}
\sigma+8\left(-1+4\lambda\right)f'\left(y\right)=0~. \ee This
equation can also be reproduced by integrating the distributional
parts of (\ref{dgmneq1}).   Substituting $f(y)$ gives the
following expression for the tension
 \be \label{btension}
\s=-2\sqrt{2}\frac{\left(1-4\lambda\right)}{\k}\sqrt{\frac{\Lambda_{5}\r}{1-4\lambda}}~.
\ee

Note that in the case, where the bulk is conformally invariant,
the brane is tensionless. Furthermore in the limit where
$\lambda\rightarrow1 $, we get the familiar result of a possitive
tension brane.

 \subsection{Curved Brane}

Instead of using a flat $g_{\m\n}$, we can introduce a metric of constant non-zero curvature. Namely, we will have

 \be
 ds^{2}_{\left(5\right)}=ds^{2}_{\left(4\right)}+dy^{2}~,
 \ee
where
 \be
 ds^{2}_{\left(4\right)}=\a\left(y\right)^{2}\left(-dt^{2}+e^{2Ht}\d_{ij}dx^{i}dx^{j}\right)
 \ee
 is a conformally flat, Einstein space of constant non-zero curvature \cite{Kogan:2000vb}. Again, eq. (\ref{dNmeq}) is automatically satisfied.
 For the  other two equations (\ref{dNeq}) and (\ref{dgmneq}), we have respectively

\bea \label{maxeq1}
-\La_{5}\r+\frac{1}{\k^{2}\a\left(y\right)^{2}}\left(6
H^{2}\k^{2}\r+\left(8-32\la\right)
\left(\a'\left(y\right)\right)^{2}\right)=0~, \eea and \bea
\label{maxeq2} \frac{3
H^{2}\r}{2\a\left(y\right)^{4}}-\frac{\La_{5}\r}{2\a\left(y\right)^{2}}-\frac{\s}{2\a\left(y\right)^{2}}\d(y)+\frac{2\left(1-4\la\right)}{\k^{2}\a\left(y\right)^{4}}\left(\left(\a'\left(y\right)\right)^{2}+\a\left(y\right)
\a''\left(y\right)\right)=0~. \eea Solving (\ref{maxeq1}) for
$H^{2}$ and substituting in (\ref{maxeq2}) we obtain \bea
\label{maxeq3} \k^{2}\a\left(y\right)\left(\La_{5}\r+2 \s
\d\left(y\right)\right)+8\left(-1+4\la\right)\a''\left(y\right)=0~.
\eea This is an equation for the warp factor $a(y)$ and its
solution will depend on the diffeomorphism breaking parameter
$\la$. It is useful to see that eqs. (\ref{maxeq1}) and
(\ref{maxeq3}) in the limit where $\la\rightarrow1$ match with
equations $(6),(7)$ of \cite{Kogan:2000vb} with
$\k^{2}\rightarrow\frac{1}{M^{3}},\r\rightarrow4M^{3}$ and
$\Lambda_{5}\rightarrow\Lambda_{5}/4M^{3}$.

 Taking into account (\ref{junc2}), we obtain as before three cases:

\subsubsection{ \underline{Case 1 $(\la<\frac{1}{4})$}}

In the case where $\la<\frac{1}{4}$, the solution is

\be \a\left(y\right)=\cos\left(m y\right)-\frac{\s
m}{\r\La_{5}}\sin\left(m|y|\right)~, \ee with \be
m^{2}=-\frac{\k^{2}\r}{8|1-4\la|}\La_{5}~, \ee  $\La_{5}$ is
negative and \be H^2={1 \over 48}\left( 8\La_5 - {\k^2 \s^2 \over
\r (1-4 \la) }\right)= {1 \over 6\, \La_5
\r^2}(\s^{2}m^{2}+\La^{2}_{5}\r^{2})~. \ee

\subsubsection{ \underline{Case 2 $(\la=\frac{1}{4})$}}

In the conformal point $\La_{5}=0$, $H^{2}=0$ and the brane is
tensionless $(\s=0)$.

\subsubsection{ \underline{Case 1 $(\la>\frac{1}{4})$}}

Finally for the case where $\la>\frac{1}{4}$ we have the following
solution

\be \a\left(y\right)=\cosh\left(m y\right)+\frac{\s
m}{\r\La_{5}}\sinh\left(m|y|\right)~, \ee  with \be
m^{2}=-\frac{\k^{2}\r}{8|1-4\la|}\La_{5}~. \label{mm}\ee  In the
above relation, $\La_{5}$ is negative and furthermore
 \bea H^{2}=\frac{4}{3}
\frac{|1-4\la|}{\r^{3}\k^{2}}
\frac{m^{2}}{\La^{2}_{5}}\left(\s^{2}m^{2}-\La^{2}_{5}\r^{2}\right)~.
\eea Summarizing, the $|H^2|$ factor for the three cases is given
by \be
\renewcommand{\arraystretch}{1.5}
|H^2|=\left\{\begin{array}{cl}\frac{4}{3} \frac{|1-4\la|}{\r^{3}\k^{2}} \frac{m^{2}}{\La^{2}_{5}}\left(\s^{2}m^{2}-\La^{2}_{5}
\r^{2}\right)&,\frac{|\La_{5}|}{m}<\frac{\s}{\r}
~~{\rm for}~dS_{4}~{\rm branes}~,\\
0&,\frac{|\La_{5}|}{m}=\frac{\s}{\r}~~{\rm for~flat~branes}~,\\
\frac{4}{3} \frac{|1-4\la|}{\r^{3}\k^{2}}
\frac{m^{2}}{\La^{2}_{5}}\left(\La^{2}_{5}\r^{2}-\s^{2}m^{2}
\right)&,\frac{|\La_{5}|}{m}>\frac{\s}{\r}~~{\rm for}~AdS_{4}~{\rm
branes}~.\end{array}\right. \ \ee For positive values of $\La_5$
we have the same solutions but for the opposite domains of
$\lambda$ and with positive sign for $m^2$ in (\ref{mm}).  The
solutions we found are similar to the solutions discussed in
\cite{Kaloper:1999sm} for curved backgrounds.

\section{Higher Extrinsic Curvature Operators}

The action (\ref{5daction2}) receives naturally radiative
corrections in the form of higher dimensional operators. The
dimension of these operators depends on $w$ if they involve
$K_{\m\n}$, or are $w$-independent if they involve only $R^{(4)}$.
It is reasonable to consider the theory only at energies that
$(R^{(4)})^{2}$ terms are subdominant, since these are bound to
introduce ghosts. For such energies, depending on the  scaling $w$
of the extra dimension, one could add higher powers of the
extrinsic curvature in the action. Since $[(R^{(4)})^{2}]=4$ and
$[K_{ij}]=w$, we need to consider these powers of K that fulfill
$n w<4$. As an example,  for $n=3$ we obtain that a value of
$w<\frac{4}{3}$ allows for cubic powers of the extrinsic curvature
to be important at the energy region where the theory is still
unitary.

Under these assumptions we are led to expand (\ref{5daction2}) by
introducing the following terms \bea \label{K3termsaction} \D
S=\int dy
dx^{4}\sqrt{-g}N\bigg(\frac{1}{\k^{4}}\frac{1}{\epsilon}\left(\a
K^{3}+\b K_{\m\n}K^{\m\n}K+\gamma
K_{\m\n}K^{\n\r}K^{\m}_{\r}\right)\bigg)~, \eea where as before
$[\k]=\frac{w-4}{2}$.  The coupling $\epsilon$ scales as
$[\epsilon]=4$ and  $\a,\b,\gamma$ are dimensionless constants.
The above action is split  into three pieces $\D S=\D S_\a + \D
S_\b + \D S_\g$  which we are going to vary separately.

The variations of the above terms with respect to $g_{\m\n}$ give
\bea \label{dgmneq2} \frac{1}{\sqrt{-g}} \frac{\d \D S_\a}{\d
g_{\m\n}}&=&N\frac{1}{\k^{4}}\frac{1}{\epsilon}\a[N\left(-K^{3}g^{\m\n}-3K^{2}K^{\m\n}\right)-\frac{3}{2}\partial_{y}
\left(K^{2}g^{\m\n}\right)-3\nabla_{\a}\left(K^{2}g^{(\m\a}N^{\n)}\right)+\frac{3}{2}N^{\a}\nabla_{\a}
\left(K^{2}g^{\m\n}\right)]\label{dgmneq3},~~\\
\frac{1}{\sqrt{-g}} \frac{\d \D S_{\b} }{\d
g_{\m\n}}&=&\frac{1}{\k^{4}}\frac{1}{\epsilon}\b[-N\left(2 K
K^{\m\s}
K^{\n}_{\,\,\s}+K^{\a\b}K_{\a\b}K^{\m\n}+K^{2}K^{\m\n}\right)\\
\nn
&-&(\partial_{y}(K\,K^{\m\n})+\frac{1}{2}\partial_{y}(K^{\a\b}K_{\a\b}g^{\m\n})+2\nabla_{\a}(K^{\a(\m}N^{\n)}K)\\
\nn
 &+&\nabla_{\a}(g^{a(\m}N^{\n)}K^{\r\s}K_{\r\s})-N^{\a}\nabla_{\a}(KK^{\m\n})-\frac{1}{2}N^{\a}\nabla_{\a}(K^{\k\l}
K_{\k\l}g^{\m\n}))] \label{dgmneq4}~,\\
\frac{1}{\sqrt{-g}} \frac{\d \D S_{\gamma} }{\d
g_{\m\n}}&=&\frac{1}{\k^{4}}\frac{1}{\epsilon}\gamma[N(\frac{1}{2}g^{\m\n}K_{\lambda\k}K^{\k\s}K_{\s}^{\,\,\lambda}
-3K_{\lambda}^{\,\,\m}K^{\n\s}K_{\s}^{\,\,\lambda}-\frac{3}{2}KK^{\m}_{\,\,\lambda}K^{\n\lambda})\\
\nn &&\qquad \qquad
\qquad-(\partial_{y}(\frac{3}{2}K^{\m}_{\,\,\s}K^{\n\s})+3\nabla_{\a}(K^{\a}_{\,\,\s}K^{(\m\s}N^{\n)})-N^{\a}
\nabla_{\a}(\frac{3}{2}K^{\m}_{\,\,\s}K^{\n\s}))]~. \eea The
variations with respect to $N_{\m}$ give\bea
\frac{1}{\sqrt{-g}} \frac{\d \D S_{\a} }{\d N_{\m}}&=&\frac{1}{\k^{4}}\frac{1}{\epsilon}\,\a\,
\left(3\, \nabla_{\n}\left(K^{2}g^{\m\n}\right)\right)\label{dNm1eq}~,\\
\frac{1}{\sqrt{-g}} \frac{\d \D S_{\b} }{\d N_{\m}}&=&\frac{1}{\k^{4}}\frac{1}{\epsilon}\b\left(\nabla_{\n}
\left(2 K^{\m\n}K+K^{\r\s}K_{\r\s}g^{\m\n}\right)\right)\label{dNm2eq}~,\\
\frac{1}{\sqrt{-g}} \frac{\d \D S_{\gamma} }{\d
N_{\m}}&=&\frac{1}{\k^{4}}\frac{1}{\epsilon}\gamma
\left(\nabla_{\n}\left(3\, K^{\m}_{\,\,\s}
K_{\s}^{\,\,\n}\right)\right)\label{dNm3eq}~. \eea Finally  the
variation with respect to $N$ is \be \label{dNeq3}
\frac{1}{\sqrt{-g}} \frac{\d \D S}{\d
N}=\frac{1}{\k^{4}}\frac{1}{\epsilon}\left(-2\right)\left(\a\,K^{3}+\b\,K_{\m\n}K^{\m\n}K+\gamma\,K_{\m\n}K^{\n\s}K_{\s}^{\,\,\m}\right).
\ee

As before we need to examine the junction conditions across the brane. Integrating (\ref{dgmneq6}) and focusing on the
 distributional part, we see that the junction condition reads

\be
\label{jcK3}
\left[\frac{2}{\kappa^{2}}\pi^{\m\n}+\frac{1}{\kappa^{4}}\frac{1}{\epsilon}\left(-\a\frac{3}{2}K^{2}g^{\m\n}
-\b\left(KK^{\m\n}+\frac{1}{2}K^{\sigma\kappa}K_{\sigma\kappa}g^{\m\n}\right)-\gamma\frac{3}{2}K^{\m}_{\,\,\sigma}K^{\n\sigma}\right)\right]^{+}_{-}
=\frac{1}{2}\sigma g^{\m\n}~. \ee

Taking into account the above variations and comparing with the
equations of motion of the quadratic terms of the extrinsic
curvature, we see that in the case where $\gamma=-4\b-16\a$ we get
exactly the same results both for flat and curved branes. This is
because in this limit (\ref{K3termsaction}) vanishes. In the case
of flat branes, even deviating from this limit, we obtain
solutions of the same form with redefined constants.

\subsection{Flat Branes}

For the case of flat branes  we consider the ansatz
(\ref{flatbrane}). Then (\ref{dNeq})
 along side with (\ref{dNeq3}) becomes
\be\label{dNeq4}
-\r\La_{5}+\frac{8\left(1-4\la\right)\,\left(f'\left(y\right)\right)^{2}}{\k^{2}}-\frac{8\left(16\a+4\b+\gamma\right)}{\epsilon\,
\k^{4}}\left(f'\left(y\right)\right)^{3}=0~. \ee Equation
(\ref{dNmeq}) together with equations (\ref{dNm1eq}),
(\ref{dNm2eq}) and (\ref{dNm3eq}) are satisfied identically.
Equation (\ref{dgmneq1}) alongside with equations
(\ref{dgmneq2}),(\ref{dgmneq3}) and (\ref{dgmneq4}) give the
following expression \bea \label{dgmneq6}
&-&\frac{\r}{2}g^{\m\n}\La_{5}
N+\frac{2}{\k^{2}}\left[\partial_{y}\p^{\m\n}+NK\p^{\m\n}+2NK^{\s\m}\p^{\n}_{\,\,\s}\right]
-\frac{1}{2}g^{\m\n}\left( \frac{2}{\k^{2}} \left(K_{\m\n}
K^{\m\n}-\la K^{2}\right)\right)-\frac{1}{2}\s
g^{\m\n}\d\left(y\right) \\ \nn
&+&\frac{1}{\k^{4}}\,\frac{1}{\epsilon}[\a\left(-K^{3}\,g^{\m\n}-3K^{2}\,K^{\m\n}-\frac{3}{2}\partial_{y}(K^{2}\,g^{\m\n})\right)\\
\nn
&+&\b\left(-2K\,K^{\m\s}\,K^{\n}_{\,\,\s}-K^{\s\k}\,K_{\s\k}\,K^{\m\n}-\partial_{y}(K\,K^{\m\n})
-\frac{1}{2}\partial_{y}(K^{\s\k}\,K_{\s\k}\,g^{\m\n})\right)\\
\nn
&+&\gamma\left(\frac{1}{2}g^{\m\n}\,K_{\s\k}\,K^{\k\lambda}\,K_{\lambda}^{\,\,\s}
-3K_{\lambda}^{\,\,\m}\,K^{\n\s}\,K_{\s}^{\,\,\lambda}-\frac{3}{2}K\,K^{\m}_{\,\,\r}\,K^{\n\r}-\frac{3}{2}\,
\partial_{y}(K^{\m}_{\,\,\s}\,K^{\n\s})\right)]=0~,
\eea
from which we get
 \bea  \label{dgmneq5}
&&\epsilon\k^{4}\Lambda_{5}\r+8\left(f'(y)\right)^{2}\left(\epsilon\k^{2}(-1+4\lambda)+(16\a+4\b+\gamma)f'(y)\right)+\epsilon\k^{4}\s\delta(y)\\
\nn
&&\qquad\qquad\qquad\qquad\qquad\qquad\qquad\qquad\qquad+2\left(2\epsilon\k^{2}(-1+4\lambda)+3(16\a+4\b+\gamma)\,f'(y)\right)f''(y)=0~.
\eea Depending on the value of the expression $(16\a+4\b+\gamma)$
we can distinguish the following cases:
\subsubsection{ \underline{Case 1 $(16\a+4\b+\gamma)=0$}}

It is clear that in the case where the dimensionless constants satisfy $(16\a+4\b+\gamma)=0$, we end with the same results as in the case of the $K^{2}$ terms.

\subsubsection{ \underline{Case 2 $(16\a+4\b+\gamma)\neq0$}}

When the above constraint is relaxed, solving (\ref{dNeq4}) for
$(f'(y))^{3}$ and substituting it to (\ref{dgmneq5}) we get \be
2\left(2\epsilon\k^{2}(-1+4\lambda)+3(16\a+4\b+\gamma)f'(y)\right)f''(y)=0~.
\ee

Now if we have that

\be
2\epsilon\k^{2}(-1+4\lambda)+3(16\a+4\b+\gamma)f'(y)=0~,
\ee
then
\bea \label{K3sol1}
 f\left(y\right)=\frac{2\, |y|\, \epsilon\,
\k^{2}\left(1-4\la\right)}{3\left(16\a+4\b+\gamma\right)}~, \eea
 and
 \bea
 \La_{5}=-\frac{32\, \epsilon^{2}\,\k^{2}\left(-1+4\la\right)^{3}}{27\,
 \left(16\a+4\b+\gamma\right)^{2}\,\r}~.
 \eea

 We note here that, contrary to the $K^{2}$ terms, the behaviour of
the warp factor $f(y)$, if it is  growing or decaying, depends on
the value of $(1-4\lambda)/(16\a+4\b+\gamma)$. Furthermore, in
this case of the $K^{3}$ terms the sign of $\Lambda_{5}$, depends
on the value of $\lambda$.

If on the other hand $f''(y)=0$, \ie \be f\left(y\right)=A y+B,
\label{K3sol2} \ee we have that
 \bea
 \La_{5}=-\frac{8A^{2}\left(\epsilon\kappa^{2}\left(-1+4\lambda\right)+\left(16\a+4\b+\gamma\right)
 A\right)}{\epsilon\kappa^{4}\r}~.
 \eea

At the conformal point where $\lambda\rightarrow1/4$, we have that
$f'(y)f''(y)=0$, so either $f(y)=const$ and $\Lambda_{5}=0$, or
$f(y)=A y+B$ and
$\Lambda_{5}=-8\left(16\a+4\b+\gamma\right)/\epsilon\kappa^{4}\r$.
Again we see that the behaviour of the warp factor and the
cosmological constant depend on the choice of parameters.

Applying in the above the junction conditions (\ref{jcK3}), we get
\be \epsilon \kappa^{4} \sigma \delta(y)+4\epsilon
\kappa^{2}(-1+4\lambda)\left[f'(y)\right]^{+}_{-}+6\left(16\a+4\b+\gamma\right)[(f'(y))^{2}]^{+}_{-}=0~.
\ee Due to the $Z_{2}$ symmetry, the quadratic term vanishes,
resulting  to the same junction equations as in the case of
$K^{2}$ terms. Substituting the solution for $f(y)$
((\ref{K3sol1})) we get \be
\sigma=\frac{16}{3}\epsilon\frac{(1-4\lambda)^{2}}{(16\a+4\b+\gamma)}~,
\ee and for the solution (\ref{K3sol2}) \be
\sigma=-\frac{8}{\kappa^{2}}(-1+4\lambda)A~. \ee Again in the
conformal limit the brane is tensionless. In order to get
significant changes to the junction conditions we have to move to
$K^{4}$ terms, since these terms will produce terms $(f'(y))^{3}$
which are even.

\section{Scalar Perturbations}

 In this section we study the scalar sector of perturbations of
the theory that has  up to quadratic terms of the extrinsic
curvature. For that purpose,  we will use the flat vacua of the
theory, analysed in section IIIA.  We consider the following
metric ansatz \footnote{Note that  the lapse function N depends
also on the  4-dimensional spacetime coordinates. Therefore, in
general,   terms depending on derivatives of N are allowed in the
action \cite{Blas:2009yd}. This would significantly increase the
number of terms in the action and to keep the analysis tractable
we assume their  couplings to be small so they can be ignored. },
\be\label{perts}
 N=e^{\a\left(x^{\n},y\right)},\quad N_{\m}=\partial_{\m}\b\left(x^{\n},y\right),\quad g_{\m\n}=e^{2\left(f\left(y\right)
 +\z\left(x^{\r},y\right)\right)} \eta_{\m\n}~,
 \ee
 which differs from the most general scalar perturbation possibly  by a perturbation of $g_{\m\n}$ of the form $2 \de_\m \de_\n E$,
   which however can be  gauged away (see \cite{Papazoglou:2009fj}).

 Using the above ansatz, we compute in Appendix A the various invariants appearing in the action.
 Inserting them back in the action and keeping terms up to quadratic order in petrurbations, we obtain the following quadratic bulk action

\be \label{pertaction}
S&=&\int\,dx^{4}dy\,e^{2f}[\r\left(3\left(\left(\partial\,\z\right)^{2}-\a\,\square^{\left(4\right)}\,\z\right)-e^{2f}
\left(1+4\z+\a+4 \a \z+\frac{\a^{2}}{2}+8\z^{2}\right)
\Lambda_{5}\right)  \\ \nn &-&\frac{2}{\k^{2}} (4 e^{2f}
(1-4\lambda)(\left(\partial_{y}f\right)^{2}+2\partial_{y}f\,\partial_{y}\z+\left(\partial_{y}\z\right)^{2}-\a\,
\left(\partial_{y}f\right)^{2}-2\a\,\partial_{y}f\,\partial_{y}\z+\frac{\a^{2}}{2}\left(\partial_{y}f\right)^{2}
\\ \nn
&+&4\z\left(\partial_{y}f\right)^{2}+8\z\partial_{y}f\,\partial_{y}\z-4\a\z\,\left(\partial_{y}f\right)^{2}+8\z^{2}\left(\partial_{y}f\right)^{2})
\\ \nn
&+&2(1-4\lambda)\partial_{y}f\a\square^{(4)}\b-4(1-4\lambda)\partial_{y}f\z\square^{(4)}\b-2(1-4\lambda)\partial_{y}\z\square^{(4)}\b\\
\nn
&-&4(1-4\lambda)\partial_{y}f\partial^{\n}\z\partial_{\n}\b+e^{-2f}(1-\lambda)
\left(\square^{(4)}\b\right)^{2})]~. \ee

  This action has one non-dynamical degree of
freedom, $\a$.  Varying the action with respect to $\a$,
produces a constraint, to be imposed to the system, which reads
\be
\label{avareq} 3\r
\square^{\left(4\right)}\,\z+\frac{4}{\k^{2}}\left(1-4\lambda\right)\partial_{y}f\,\square^{\left(4\right)}\b&=&\frac{8}{\k^{2}}e^{2f}\left(1-4\lambda\right)\left(\left(\partial_{y}f\right)^{2}+2\partial_{y}f\partial_{y}\z-\a\left(\partial_{y}f\right)^{2}+4\z\left(\partial_{y}f\right)^{2}\right)\\
\nn & &-e^{2f}\r\Lambda_{5} \left(1+4\z+\a\right)~. \ee

Using equation (\ref{avareq}) we can eliminate
$\a$ in favor of the $\z$ and $\b$  in the action
(\ref{pertaction}) and then we obtain

\be \label{zetaaction}
S&=&\int\,dx^{4}dy\,e^{2f}[24\partial_{y}f(\partial_{y}\z)\square^{\left(4\right)}\z\frac{\left(-1+4\lambda\right)}
{\Lambda_{5}\k^{2}}+16\partial_{y}f\partial_{y}\z\left(-1+4\lambda\right)e^{2f}\frac{(1+4\z)}{\k^{2}}\\
\nn
&+&3\r\left(\partial\z\right)^{2}+\frac{9\r}{4\Lambda_{5}}e^{-2f}\left(\square^{\left(4\right)} \z\right)^{2}-\frac{3}{2 \k^{2}}e^{-2f}\left(\square^{\left(4\right)} \b\right)^{2}-6(\partial_{y}f) e^{-2f}\frac{(-1+4\lambda)}{\k^{2}\Lambda_{5}} \square^{\left(4\right)} \b \square^{\left(4\right)} \z
-2e^{2f}\left(1+4\z+8\z^{2}\right)\Lambda_{5}\r]
\\ \nn
&-&\int\,dx^{4}dy\,e^{2f}\left(1+4\z+8\z^{2}\right)e^{2f}\s\delta(y)~.
\ee The last term in the above action is the brane boundary term
appearing in (\ref{5daction2}).
The above action, as it is explained in the Appentix A, after appropriate partial integrations (while assuming appropriate boundary conditions), can be brought to the final
form \be \label{result}
S&=&\int\,dx^{4}dy\,e^{2f}[\frac{9\r}{4\Lambda_{5}}e^{-2f}\left(\square^{\left(4\right)} \z\right)^{2}-\frac{3}{2 \k^{2}}e^{-2f}\left(\square^{\left(4\right)} \b\right)^{2}-6(\partial_{y}f) e^{-2f}\frac{(-1+4\lambda)}{\k^{2}\Lambda_{5}} \square^{\left(4\right)} \b \square^{\left(4\right)} \z  \\ \nn
& &\qquad \qquad \quad +\frac{6\sqrt{2}}{\k}\frac{(1-4\lambda)}{\Lambda_{5}}\sqrt{\frac{\Lambda_{5}\r}{1-4\lambda}}\left(\partial\,\z\right)^{2}\,\delta\left(y\right)]~.
\ee

This is the main result of our work. The terms in the first line
are higher derivative terms for the two  dynamical degrees of
freedom $\zeta$ and $\b$, which have kinetic mixing between them.
As a check of the correctness of the above action, one can vary it
with respect to $\b$. Then we obtain the same result with the
variation of (\ref{pertaction}) which reads \be \label{bvareq}
\left(1-\lambda\right)e^{-2f}\square^{\left(4\right)}\b=\left(1-4\lambda\right)\left(\partial_{y}\z-\partial_{y}f
\a\right)~. \ee The equivalence can be seen by substituting
(\ref{avareq}) in (\ref{bvareq}). Similarly, the variation with
respect to $\zeta$ gives the same result once the constraint is
taken into account.

The action (\ref{result}) has certain characteristics. First, all
the bulk terms involve four derivatives of  brane coordinates, and
certain terms have four time derivatives. Secondly, the brane term
is a ghost-like kinetic term for $\zeta$. Once the wave-functions
for the  two dynamical modes have been substituted, the four
dimensional action will be consisted from the ghost kinetic term
for $\zeta$, plus higher derivative terms. These terms will appear
in the action multiplied with different scales. However, whatever
the hierarchy of these scales may be, there will always be some
ghost problem in the spectrum: either from the quadratic kinetic
term, or from the higher derivative terms.

These modes will be present even after the restoration of the
five-dimensional diffeomorphism symmetry, \ie when $\lambda \to
1$. This last characteristic, is very similar to what happens in
massive gravity, namely the vDVZ discontinuity
\cite{vanDam:1970vg,Zakharov:1970cc}. In that case, similarly the
longitudinal mode of the massive graviton does not decouple in the
limit of vanishing Pauli-Fierz mass. In our example, we have some
even worse result, since the remaining modes have ghost behaviour.
These problems are probably due to the explicit breaking of the
diffeomorphism invariance in the theory. The theory may not appear
problematic at the level of background solutions, but nevertheless
these   problems reveal themselves once the theory is  perturbed.

\section{Conclusions}

We investigated the consequences of an assumption that the  fifth
extra dimension scales differently than the other four dimensions.
To achieve this we considered a five-dimensional theory with a
cosmological constant and a three brane embedded in it. In this
theory,  the full five-dimensional diffeomorphism group is
explicitly broken down to its foliation-preserving
four-dimensional subgroup. The foliation we used  involves an
extra space dimension and therefore the four-dimensional Lorentz
invariance is intact. Because of the different scaling of the
extra spacial dimension, higher powers of the extrinsic curvature
of the four-dimensional hypersurfaces are allowed in the theory up
to the energies  that ghost-bearing higher order dimensional
operators appear.

We made a systematic study of the local solutions of this theory.
For maximally symmetric backgrounds and up to second order in
extrinsic curvature we found all solutions for flat and curved
branes. These solutions are similar to the previously obtained
solutions for flat and curved branes, they are however
characterized by a parameter $\lambda$ that expresses the breaking
of the five-dimensional diffeomorphism invariance. We also
obtained solutions in the flat brane limit by including higher
order in extrinsic curvature terms. These solutions except their
dependence on $\lambda$ also depend on the coefficients by which
the higher order  extrinsic curvature terms enter in the action.

Having explicitly broken the Lorentz invariance in five
dimensions, we looked for possible effects on the four dimensional
spacetime. We performed a scalar perturbations analysis of the
theory for up to quadratic terms in the extrinsic curvature. We found that, as a result of the breaking of the
diffeomorphism invariance, two dynamical scalar degrees of freedom, appear.
 Both of them are characterized by the fact that they come as higher derivative corrections to the action.  Moreover one of these modes
 appears with a ghost-like brane kinetic term. These pathologies can be attributed to the explicit breaking of diffeomrphism
 invariance along the extra space dimension.


\textbf{Acknowledgments:} M.T. would like to thank the Institute of Cosmology  and Gravitation of Portsmouth for kind hospitality during the first stages of the work. The authors would like to thank Antonio Padilla for useful discussions.

\begin{appendix}

\section{Technical Details for Scalar perturbations}

In this Appendix we give the technical details for deriving the
action (\ref{result}) in section V. We start with the generalize
action (\ref{5daction2}) ignoring  for the moment the boundary
brane term
 \be\label{A5daction2pert}
 S=\int N \sqrt{-g}\, dy\, dx^{4}\left[\frac{\r}{2}\left(R^{\left(4\right)}-2\La_{5}\right)
 -\frac{2}{\k^{2}} \left(K_{\m\n} K^{\m\n}-\la K^{2}\right)\right]~.\ee
We perturb the above action, in the flat brane limit discussed in
section IIIA, using the perturbation ansatz (\ref{perts}) for the
scalar perturbations. The extrinsic curvature and its trace is
given by
 \be
K^{\m\n}&=& e^{-4\left(f+\z+\frac{\a}{4}\right)}\left[
e^{2\left(f+\z\right)}\partial_{y}\left(f+\z\right)\eta^{\m\n}-\partial^{\m}\partial^{\n}
\beta+\partial^{\m}\z\partial^{\n}\b+\partial^{\n}\z\partial^{\m}\b-\eta^{\m\n}\partial^{\lambda}\z\partial_{\lambda}\b\right]~,\\
K&=&e^{-2\left(f+\z+\frac{\a}{2}\right)}\left[4\,e^{2\left(f+\z\right)}\partial_{y}\left(f+\z\right)
-\square^{\left(4\right)}\b-2\partial^{\lambda}\z\partial_{\lambda}\b\right]~,
\ee
 where
$\square^{\left(4\right)}=\eta^{\m\n}\partial_{\m}\partial_{\n}$, while the Ricci scalar is \be
R=-6\,e^{-2\left(f+\z\right)}\left(\square^{\left(4\right)}\z+\left(\partial\,\z\right)^{2}\right)~.
\ee

Collecting the above
terms, the perturbed action, up to quadratic order, reads  \be
\label{AAfullaction}
S&=&\int\,dx^{4}dy\,e^{2f}[\r\left(3\left(\left(\partial\,\z\right)^{2}-\a\,\square^{\left(4\right)}\,\z\right)-e^{2f}
\left(1+4\z+\a+4 \a \z+\frac{\a^{2}}{2}+8\z^{2}\right)
\Lambda_{5}\right)  \\ \nn &-&\frac{2}{\k^{2}} (4 e^{2f}
(1-4\lambda)(\left(\partial_{y}f\right)^{2}+2\partial_{y}f\,\partial_{y}\z+\left(\partial_{y}\z\right)^{2}-\a\,
\left(\partial_{y}f\right)^{2}-2\a\,\partial_{y}f\,\partial_{y}\z+\frac{\a^{2}}{2}\left(\partial_{y}f\right)^{2}
\\ \nn
&+&4\z\left(\partial_{y}f\right)^{2}+8\z\partial_{y}f\,\partial_{y}\z-4\a\z\,\left(\partial_{y}f\right)^{2}+8\z^{2}\left(\partial_{y}f\right)^{2})
\\ \nn
&+&2(1-4\lambda)\partial_{y}f\a\square^{(4)}\b-4(1-4\lambda)\partial_{y}f\z\square^{(4)}\b-2(1-4\lambda)\partial_{y}\z\square^{(4)}\b\\
\nn
&-&4(1-4\lambda)\partial_{y}f\partial^{\n}\z\partial_{\n}\b+e^{-2f}(1-\lambda)
\left(\square^{(4)}\b\right)^{2})]~. \ee The above action contains
one non-dynamical field, namely $\a$. Varying the action with
respect to $\a$, we get a constraint,
\be
\label{Aaavareq} 3\r
\square^{\left(4\right)}\,\z+\frac{4}{\k^{2}}\left(1-4\lambda\right)\partial_{y}f\,\square^{\left(4\right)}\b&=&\frac{8}{\k^{2}}e^{2f}\left(1-4\lambda\right)\left(\left(\partial_{y}f\right)^{2}+2\partial_{y}f\partial_{y}\z-\a\left(\partial_{y}f\right)^{2}+4\z\left(\partial_{y}f\right)^{2}\right)\\
\nn & &-e^{2f}\r\Lambda_{5} \left(1+4\z+\a\right)~, \ee
while the equation for $\b$ is the following
\be \label{Abvareq}
\left(1-\lambda\right)e^{-2f}\square^{\left(4\right)}\b=\left(1-4\lambda\right)\left(\partial_{y}\z-\partial_{y}f
\a\right)~.\ee
Using  (\ref{Aaavareq}) we can eliminate $\a$
 from the  action (\ref{AAfullaction}) and get the
following expression for the action where we have two dynamical fields appearing in the action, $\z$ and $\b$
\be \label{Azetaaction}
S&=&\int\,dx^{4}dy\,e^{2f}[24\partial_{y}f(\partial_{y}\z)\square^{\left(4\right)}\z\frac{\left(-1+4\lambda\right)}
{\Lambda_{5}\k^{2}}+16\partial_{y}f\partial_{y}\z\left(-1+4\lambda\right)e^{2f}\frac{(1+4\z)}{\k^{2}}\\
\nn
&+&3\r\left(\partial\z\right)^{2}+\frac{9\r}{4\Lambda_{5}}e^{-2f}\left(\square \z\right)^{2}-\frac{3}{2 \k^{2}}e^{-2f}\left(\square \b\right)^{2}-6(\partial_{y}f) e^{-2f}\frac{(-1+4\lambda)}{\k^{2}\Lambda_{5}} \square \b \square \z
-2e^{2f}\left(1+4\z+8\z^{2}\right)\Lambda_{5}\r]
\\ \nn
&-&\int\,dx^{4}dy\,e^{2f}\left(1+4\z+8\z^{2}\right)e^{2f}\s\delta(y)~.
\ee
Note that we have restated  the boundary brane term of the
original action which we had neglected so far. This action can be
simplified as follows. Consider the following terms  \be \int
dx^{4}dy\,e^{2f}[16\partial_{y}f\partial_{y}\z\left(-1+4\lambda\right)e^{2f}\frac{1+4\z}{\k^{2}}-2e^{2f}\left(1+4\z+8\z^{2}\right)
\Lambda_{5}\r]~, \ee
and perform  a partial integration with respect to the extra
dimension    using (\ref{warpfact1}). Then these terms are equal
to \be -\int
dx^{4}dy\,e^{4f}2\sqrt{2}\frac{\left(1-4\lambda\right)}{\kappa}\sqrt{\frac{\Lambda_{5}\r}{1-4\lambda}} \left(1+4\z+8\z^{2}\right)\delta(y)~,\nn
\ee which is exactly the same as the contribution of the boundary
term appearing in (\ref{Azetaaction})  with an opposite sign,
after having used  the value of the brane tension
(\ref{btension}). From the remaining terms of the action
(\ref{Azetaaction}) consider the following terms   \be
\int\,dx^{4}dy\,e^{2f}[24\partial_{y}f(\partial_{y}\z)\square^{\left(4\right)}\z\frac{\left(-1+4\lambda\right)}
{\Lambda_{5}\k^{2}}+3\r\left(\partial\z\right)^{2}]~. \ee

Observing that  \be \int
dx^{4}dy\,e^{2f}\partial_{y}[\partial_{\m}\z\partial^{\m}\z]=\int
dx^{4}dy\,e^{2f}\left(-2\left(\partial_{y}\z\right)\square^{\left(4\right)}\z\right),
\ee where we have performed a partial integration with respect to
the brane coordinates and assumed appropriate boundary conditions,
the above terms  become \be \label{midterm}
\int\,dx^{4}dy[3\r\left(\partial\z\right)^{2}e^{2f}+12\left(-2(\partial_{y}\z)\square^{\left(4\right)}\z\right)\partial_{y}f\,e^{2f}\frac{\left(1-4\lambda\right)}{\Lambda_{5}\k^{2}}] &=&\nn\\
\int\,dx^{4}dy[3\r\left(\partial\z\right)^{2}e^{2f}+12\partial_{y}\left(\left(\partial\,\z\right)^{2}\right)\partial_{y}f\,e^{2f}\frac{\left(1-4\lambda\right)}{\Lambda_{5}\k^{2}}] &=&\nn\\
\int\,dx^{4}dy\,e^{2f}
\frac{6\sqrt{2}}{\k}\frac{(1-4\lambda)}{\Lambda_{5}}\sqrt{\frac{\Lambda_{5}\r}{1-4\lambda}}\left(\partial\,\z\right)^{2}\,\delta\left(y\right)~.
\ee Using (\ref{midterm}) the final result of the perturbed action
(\ref{5daction2})  under the scalar perturbations of the form
(\ref{perts}) is \be
S&=&\int\,dx^{4}dy\,e^{2f}[\frac{9\r}{4\Lambda_{5}}e^{-2f}\left(\square \z\right)^{2}-\frac{3}{2 \k^{2}}e^{-2f}\left(\square \b\right)^{2}-6(\partial_{y}f) e^{-2f}\frac{(-1+4\lambda)}{\k^{2}\Lambda_{5}} \square \b \square \z  \\ \nn
& &\qquad \qquad \quad +\frac{6\sqrt{2}}{\k}\frac{(1-4\lambda)}{\Lambda_{5}}\sqrt{\frac{\Lambda_{5}\r}{1-4\lambda}}\left(\partial\,\z\right)^{2}\,\delta\left(y\right)]~.
\ee

\end{appendix}

\end{document}